\documentstyle[times,aps,prl,epsfig,balanced]{revtex}

\newcommand{\BE}{\begin{equation}}
\newcommand{\EE}{\end{equation}}
\newcommand{\BA}{\begin{eqnarray}}
\newcommand{\EA}{\end{eqnarray}}
\newcommand{\bc}{\begin{center}}
\newcommand{\ec}{\end{center}}
\newcommand{\bi}{\begin{itemize}}
\newcommand{\ei}{\end{itemize}}
\newcommand{\ie}{{\it i.e. }}
\newcommand{\eg}{{\it e.g. }}
\newcommand{\ignore}[1]{}

\begin{document}

\draft

\title{Transmission of Information and Herd Behavior: an Application to
Financial Markets}

\author{V\'{\i}ctor M. Egu\'{\i}luz$^{1,2,}$\cite{email},
Mart\'{\i}n G. Zimmermann$^{1,3,}$\cite{ba}}

\address{
$^1$Instituto Mediterr\'aneo de Estudios Avanzados IMEDEA
\cite{imedea}(CSIC-UIB), E-07071 Palma de Mallorca (Spain)
\\$^2$Center for Chaos and Turbulence Studies, The Niels Bohr Institute,
Blegdamsvej 17, DK-2100 Copenhagen \O \ (Denmark)
\\$^3$Departamento de F\'{\i}sica, Fac. Ciencias Exactas y
Naturales, Universidad de Buenos Aires,
Buenos Aires (Argentina) }

\date{\today}

\maketitle

\begin{abstract}
We propose a model for stochastic formation of opinion clusters,
modelled by an evolving network, and herd behavior to account for
the observed fat-tail distribution in returns of financial-price
data. The only parameter of the model is $h$, the rate of
information dispersion per trade, which is a measure of herding
behavior. For $h$ below a critical $h^*$ the system displays a
power-law distribution of the returns with exponential cut-off.
However for $h>h^*$ an increase in the probability of large
returns is found, and may be associated to the occurrence of
large crashes.
\end{abstract}

\pacs{PACS numbers: 87.23.Ge, 02.50.Le, 05.45.Tp, 05.65.+b}

\begin{twocolumns}
Recently, there has been a significant interest in applications of
physical methods in social and economical sciences
\cite{Oliveira99}.
In particular, the analysis of financial stock market prices have
been found to exhibit some universal characteristics similar to
those observed in physical systems with large number of
interacting units, and several microscopic models have been
developed to study them
\cite{Mantegna99,Bouchaud97,Lux99}.
For example the distribution of the so-called {\em returns}, \ie
the logarithmic change of the market price, has been observed to
present pronounced tails larger than in a Gaussian distribution
\cite{Mantegna99,Bouchaud97,Mantegna95,Lux96,Gopikrishnan99}.
Several models have been put forward which phenomenologically
show the fat-tail distributions. Among the more sophisticated
approaches are dynamic multi-agent models\cite{Lux99,Bak97} based
on the interaction of two distinct agents populations, (``noisy''
and ``fundamentalists'' traders) which reproduces the desired
distributions, but fails to account for the origin of the
universal behavior.
An alternative approach, explored in this Letter,
is that {\em herd behavior} \cite{Topol91,Bannerjee93}
may be sufficient to induce the desired
distributions. Herding assumes some degree of coordination
between a group of agents. This coordination may arise in
different ways, either because agents share the same
information, or they follow the same rumor. This approach has
been recently formalized by Cont and Bouchaud\cite{Cont97}, as a
{\em static} percolation model.

We present a model for opinion cluster formation and information
dispersal by agents in a network. As a first approach to model this
complicated social behavior we consider: (i) a random dispersion of
information, (ii) agents sharing the {\em same} information form a
group that make decisions as a whole (herding), and (iii) whenever a
group performs an action,  the network necessarily adapts to this
change.  We then apply the model to study the price dynamics in a
financial market.
Our results show that when the information
dispersion is much faster than trading activity, the distribution of
the number of agents sharing the same information behaves as a
power-law. Using a linear relationships for the price update in terms
of the order size \cite{Cont97,Farmer98}, the price-returns also
exhibit this universal feature (with a different exponent). On the
other hand when the dispersion of information becomes slower, a smooth
transition to truncated exponential tails is observed, with a portion
of the distribution remaining close to the power law. In our approach
the average connectivity $c(t)$ of the network is driven by the rumors
in a {\em dynamic} way, and provides an extension to the static
percolation model proposed by Cont and Bouchaud\cite{Cont97}, where the
average connectivity was a fixed external parameter. We find that the
fat-tails distributions are observed  even when the time-average
density $c(t)$ is far from the critical threshold $c^*=1$ found in
Ref.~\cite{Cont97}.

{\em The model.}
We consider a system composed by $N$ agents,
represented by vertices in a network. The state of agent $l$ is
represented by $\phi_l=\{0, +1, -1\}$ corresponding
to an inactive state (waiting [$\phi_l= 0]$), and two active
states (either buying [$\phi_l= +1$] or selling [$\phi_l= -1$]).
Agents can be isolated or
connected through links forming a {\em cluster},
\ie
those who share the same information.
Initially, all
agents are inactive ($\phi_l= 0, \forall l$) and isolated (\ie no
links between them). The network of links evolves dynamically in
the following way. At each time step $t_i$:
\begin{enumerate}
\item an agent $j$ is selected at random,
\item with probability $a$, the state of $j$ becomes active by
randomly choosing the state $1$ or $-1$, and {\em instantly} all agents belonging to the same
cluster follow this same action by imitation. The aggregate state of
the system $s_i=s (t_i) = \sum_{l=1,N} \phi_l$ and the total size
of the cluster $|s_i|$ are computed. After that the cluster is
broken up into isolated agents, removing all links inside the
cluster, and resetting their state $\phi_l=0, \forall l$;
\item with probability $(1-a)$, the state of $j$ remains inactive
($\phi_j=0$), and instead, a new
link between agent $j$ and any other agents from the whole network is
established.
\end{enumerate}
This process is repeated from step (1).

The evolution of the system is characterized by a succession of
discrete events $s_1, s_2, \ldots $, which correspond
to avalanches, occurring instantly whenever an activation occurs.
Interspeded between these events, new links are incorporated to
the network.
A quantity relevant in bond percolation is the 
connectivity of the network, $c_i=c(t_i)$, defined as the average
number of links per agent, which will grow as long as the agents remain
inactive, or decay when an avalanche occurs. 
Thus, we expect that the system will reach an asymptotic regime
where the connectivity fluctuates around a mean value, which
increases as the activation rate $a$ decreases.


{\em An application to price dynamics.}
Consider now that the above agents participate in a stock market. When an agent
becomes active, a buy ($\phi_j=+1$) or a sell ($\phi_j=-1$) order is posted to
an external centralized market-maker. When inactive (\ie waiting), the agent
disperses an information unit represented by the random addition of a link to
the cluster. All the members in the cluster share the same information, thus it
constitutes a group of opinion or an information cluster. This process includes
the possibility of cluster merging, in which case the information is shared
among the new set.

Herding assumes that agents are not making decisions independently, but that
each agent acts as belonging to a group that makes a collective action. In the
above model the herding behavior is represented by the instantaneous imitation
of an activated agent throughout the information cluster. We notice also that
the instantaneous imitation process we apply above may also result when
financial agents all use similar tools for analysis (and similar know-how).
Also, we assume that after an activation
event takes place the information content of the cluster is no longer
useful, so all links in the cluster are removed after the 
order has been placed. 

The parameter $a$ controls the rate of trading activity vs.
information dispersion, and appears as the only adjustable
parameter of the model. For $a \to 1$ only trading activity takes
place. Thus, starting with some randomly dispersed links, the
evolution of the market will asymptotically approach that of
isolated agents trading in the market, without large clusters and
thus no herding behavior. On the other hand, for small $a \ll 1$,
dispersion of information occurs on most time steps,
increasing the internal connectivity.
Initially the empty network has time to build many clusters, which
eventually merge into bigger clusters, until most agents belong to
a super-cluster. When an order arrives, this will most probably
come from the agents in the super-cluster, inducing a large impact
on the market. Although an extreme scenario, we can estimate that
this should occur when $a\ll O(1/N)$. From the above discussion we
can define the
parameter $h\equiv 1/a -1$ as the ``herding parameter": no herding
occurs ($h=0$) for $a=1$, while herding is observed ($h>0$) for
$a<1$. Alternatively this parameter also tells how many links are born
between two trade orders, \ie the rate of information dispersion.

Finally we introduce the price index dynamics, executed by  an
external centralized market-maker. 
We follow the simple update rule for the price index $P$
discussed in Ref.~\cite{Farmer98}, which arises considering
that  each order  acts as an ``impact" to the price proportional
to the size of it.  In our case, when activity takes place at
step $i$, all agents in the expiring cluster act simultaneously,
so the size of the order is $|s_i|$. Therefore we consider that
$P$ evolves as  $P(t_{i+1})= P(t_i) \exp(s_i/\lambda)$, where 
$\lambda$ is a parameter  which
controls the size of the updates and provides a measure of the
liquidity of the market. With the above rule, the price return
$R(t_i) = \ln (P(t_i)) - \ln(P(t_{i-1}))$ is proportional to the order
size. 
Other nonlinear suggestions exist for the price
update \cite{Zhang99}, which will modify the exponents of the
distribution of returns. However we stress that the power-law
features observed in this model persist with this modification and
are a consequence of the network growth and annihilation of links.

We have performed numerical simulations for a population of
$N=10^4$ agents and for different values of the herding
parameter. In the following simulations the time unit has
been rescaled to that of the average time to place an order:
$t^*=t/a$. For example, a value of $a=0.01$ [$h=99$] corresponds to
an average of a buy or sell order every $100$ iterations, or in
other words the ``market time" $t^*$ of one unit will correspond
in average to $99$ agents dispersing a rumor and one (buy or
sell) order. Figure \ref{findex}(a) displays a typical
evolution of the market price $P(t)$, Fig.~\ref{findex}b shows
the corresponding returns $R(t)$, while the evolution of the 
connectivity $c(t)$ is shown in Fig.~\ref{findex}(c). The
latter panel displays the connectivity fluctuates around the time
average $\langle c \rangle=0.78$, with some fluctuations overshooting the
critical value $c^*=1$. The mean value of the connectivity and its
standard deviation increases 
with increasing herding
parameter $h$ (decreasing $a$).

In Fig.~\ref{fret} we show the distribution of returns for three
different herding parameters $h = 2.33$, $9$ and $99$ 
[$a = 0.30$, $0.10$ and $0.01$, respectively]. The solid line shows a
power-law $R^{-\alpha}$ with exponent $\alpha=1.5$. Note that in
all cases one observes power-law decay in a range of returns.
Moreover, for increasing $h$ this range increases, up to a
critical value $h^*$ where we conjecture a power-law will be
fitted on the whole range. For $h<h^*$ the distributions display a
continuous crossover to an exponential cut-off, where the time
average $c(t)$ is far from $c^*$. However for $h>h^*$, the
time-average $c(t)$ lies very close to the critical threshold
$c^*$, and the distribution changes qualitatively. A relative
increase in the probability of extremely high returns is observed,
which would favor the creation of ``financial crashes".  We remark
that this bump in the distribution has not been reported using
financial time series, but is common in other physical systems
\cite{Corralm99}. In this regime clusters of the system size are
created and produce the large returns.

The distribution of returns is related in this model to the
distribution of cluster. In fact, if $\beta$ is the exponent for
the distribution of cluster sizes and $\alpha$ is the exponent for
the distribution of returns, then the distribution of returns is
equal to the distribution of cluster times the probability to
chose a given cluster that is proportional to its size:
$\mbox{prob}(R) \approx R^{-\alpha} \approx s s^{-\beta}$. The
exponents are related by $\alpha = \beta - 1$. We plot in
Fig.~\ref{fd} the averaged distribution of clusters. The solid
line represents a power law with exponent $\beta = 2.5$. This
result agrees with the previous calculation and with theoretical
results on stationary random graphs that predict an exponent of
$5/2$ at the critical point \cite{Cont97}. Recently, this exponent
was found by D'Hulst and Rodgers \cite{DHulst99} in a mean-field
analysis of our model. They also extended the model by allowing
multiple rumors to be dispersed at a single time-step, finding the
exponent is robust.

In Fig.~\ref{ft} we show a linear-log plot of the probability
distribution of normalized returns, defined as $(R-\langle R
\rangle)/\sigma$ with the average return $\langle R \rangle$
(about 0) over the time series and the volatility $\sigma =
(\langle R^2 \rangle - \langle R \rangle^2)^{1/2}$, for $a=0.10$
and different time intervals $\Delta t = 1, 10, 100, 1000$. With
an increasing time interval $\Delta t$, a crossover towards a
Gaussian distribution is observed from the figure, in agreement
with empirical financial data \cite{Gopikrishnan99}.

In summary, we have presented a self-organized model for the
propagation of information and the formation of groups and we have
applied it to the description of herd behavior in a financial
market.
We suppose that the propagation of information within the network
follows a random process, and the traders can be classified into
groups (clusters) having the same opinion. In our description the
size and number of clusters evolves in time reflecting the
information content of the market. This is controlled by the
herding parameter, which is a measure of the rate of rumor
dispersion.
Our numerical calculations show that for herding behavior below a
critical value $h<h^*$, herding produces qualitatively the same
distributions observed in empirical data: a power-law range and an
exponential cut-off. However for sufficiently high herding
parameter $h>h^*$, a qualitatively change is observed, where the
probability of large crashes {\em increases}. We conclude that
information dispersion and herding may be able to account for the
occurrence of crashes. In our approach, we propose a mechanism for
the fluctuation of the connectivity of the network in contrast
with fixed \cite{Cont97} or random ``sweeping" of the connectivity
in other percolation-like models \cite{Cont97,Stauffer99a}. More
elaborate mechanisms for the activation of the agents, for opinion
conflicts when two cluster merge, and feedback between the price
index and the activation should be incorporated in the model to
make it more realistic. Slight modifications of the model could be
applied to study, for example, social systems of opinion
formation.

We acknowledge fruitful discussions with Dante R. Chialvo, Juan
Pedro Garrahan, P. Gopikrishnan and Emilio Hern\'andez-Garc\'{\i}a. Financial
support has been provided by PB94-1167 and PB97-0141-C02-01
projects from Direcci\'on General de Investigaci\'on
Cient\'{\i}fica y T\'ecnica (DGICYT, Spain).

\begin{figure}[t]
\centerline{\epsfig{file=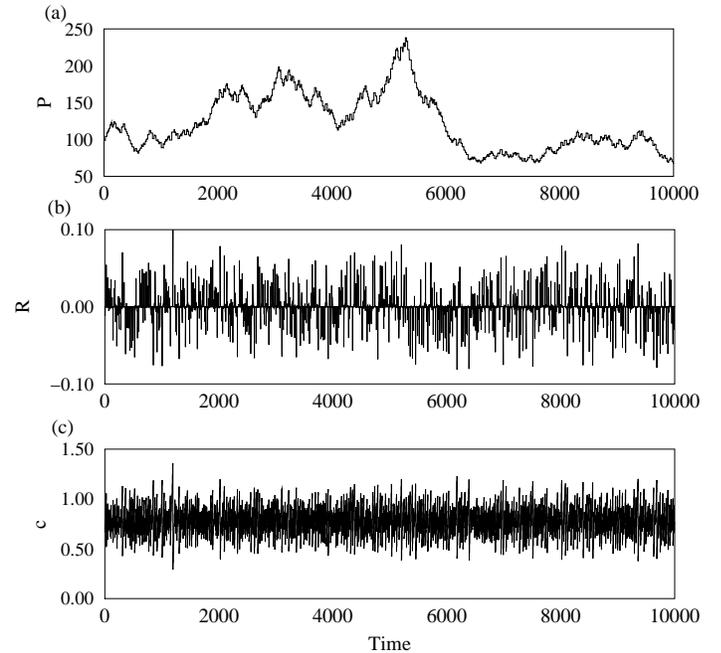,width=.5\textwidth,angle=0}}
\caption{ \small{(a) Time series of the typical evolution of the
market price $P(t)$, (b) the corresponding returns $R(t) = \ln
P(t) - \ln P(t-1)$, and (c) the connectivity $c(t)$. The mean
value of the connectivity is $\langle c \rangle=0.78$ and the
standard deviation $\sigma=0.14$. Number of agents $N=10^4$, $h = 99$
[$a=0.01$] and liquidity $\lambda=5 \times 10^{4}$. }} \label{findex}
\end{figure}

\begin{figure}[t]
\centerline{
\epsfig{file=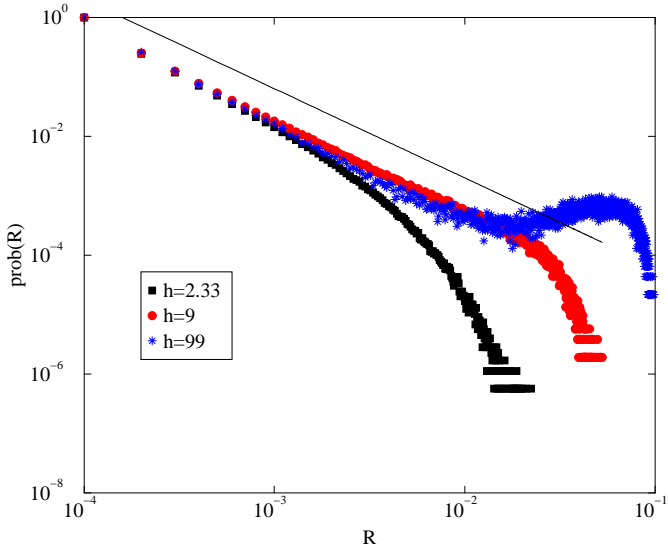,width=.5\textwidth,angle=0}}
\caption{\label{fret} \small{Log-log plot of the distribution (in
arbitrary units) of returns $R$ (in arbritary units)
for different herding parameters $h = 2.33$, $9$, $99$
[equivalently $a= 0.30$, $0.10$, $0.01$].
The solid line shows a power-law $R^{-\alpha}$ with exponent
$\alpha=1.5$. The total time integration was of $t^*\sim
10^6-10^8$ units. }}
\end{figure}

\begin{figure}[t]
\centerline{
\epsfig{file=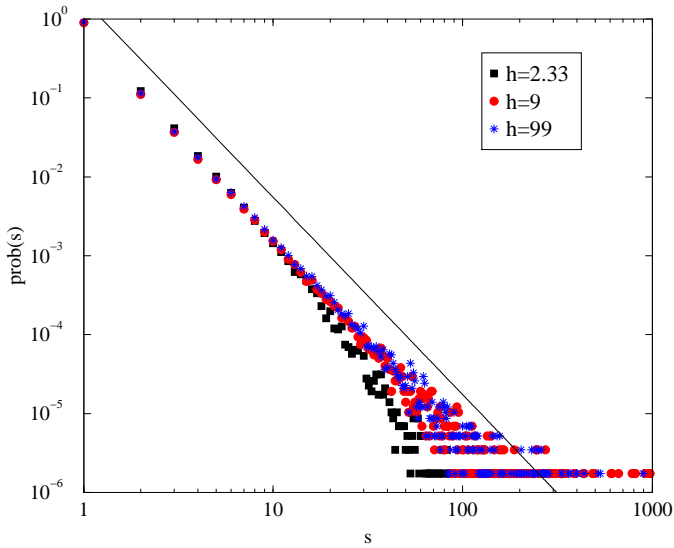,width=.5\textwidth,angle=0}}
\caption{\label{fd} \small{Log-log plot of the averaged
distribution of cluster sizes $|s|$ for $h=2.33$, $9$, $99$
[$a= 0.30$, $0.10$, $0.01$]. Solid line
shows a power law $|s|^{-\beta}$ with exponent $\beta=2.5$, and for
a total time integration $t^*\sim 10^8$ units.}}
\end{figure}

\begin{figure}[t]
\centerline{ \epsfig{file=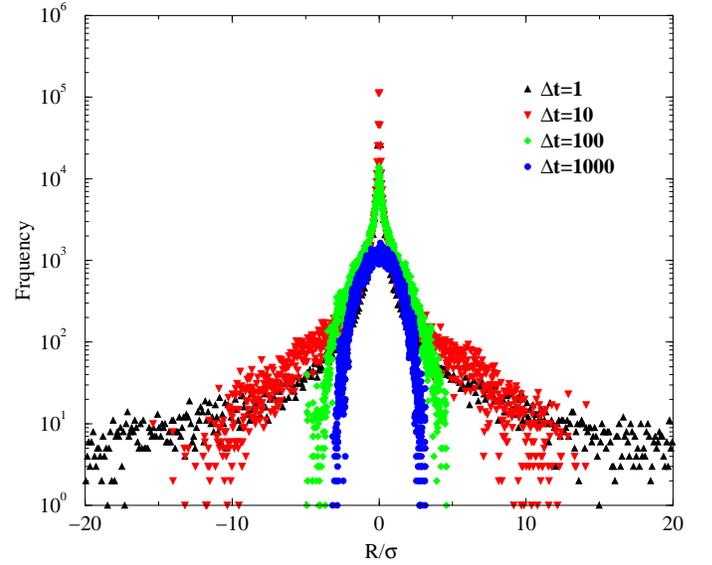, width=.5\textwidth,angle=0}}
\caption{\label{ft} \small{Semi-log plot of the distribution of
the normalized returns ($R/\sigma$, where $\sigma$ is the standard
deviation) for different time intervals $\Delta t =1$, $10$,
$100$, and $1000$. Parameter value $h=99$ [$a = 0.01$]. A crossover toward
a Gaussian distribution is shown with the increasing of time
interval.}}
\end{figure}

\end{twocolumns}
\end{document}